\documentclass[preprint]{ptephy_v1}

\preprintnumber{KYUSHU-HET-259, OU-HET-1184}

\usepackage{amsmath,amssymb,amsthm}
\usepackage{mathtools}

\numberwithin{equation}{section}
\usepackage{physics}

\newtheorem*{thm*}{Theorem}
\newtheorem*{lem*}{Lemma}
\newtheorem*{cor*}{Corollary}

\usepackage{graphicx,xcolor}
\usepackage[compat=1.1.0]{tikz-feynhand}
\usetikzlibrary{patterns}
\usetikzlibrary{decorations.markings}

\usepackage{hyperref}

\title{Note on lattice description of generalized symmetries in $SU(N)/\mathbb{Z}_N$ gauge theories}
\author{Motokazu Abe\thanks{abe.motokazu@phys.kyushu-u.ac.jp}}
\affil{Department of Physics, Kyushu University,
744 Motooka, Nishi-ku, Fukuoka 819-0395, Japan}
\author{Okuto Morikawa\thanks{o-morikawa@het.phys.sci.osaka-u.ac.jp}}
\affil{Department of Physics, Osaka University,
Toyonaka, Osaka 560-0043, Japan}
\author[1]{Soma Onoda}

\begin{document}
\begin{abstract}
 Topology and generalized symmetries in the $SU(N)/\mathbb{Z}_N$ gauge theory
 are considered in the continuum and the lattice.
 Starting from the $SU(N)$ gauge theory with the 't~Hooft twisted boundary condition,
 we give a simpler explanation
 of the van~Baal's proof on the fractionality of the topological charge.
 This description is applicable to both continuum and lattice
 by using the generalized L\"uscher's construction of topology on the lattice.
 Thus we can recover the $SU(N)/\mathbb{Z}_N$ principal bundle from lattice $SU(N)$ gauge fields
 being subject to the $\mathbb{Z}_N$-relaxed cocycle condition.
 We explicitly demonstrate the fractional topological charge,
 and verify an equivalence with other constructions reported recently based on different ideas.
 Gauging the $\mathbb{Z}_N$ $1$-form center symmetry enables lattice gauge theories
 to couple with the $\mathbb{Z}_N$ $2$-form gauge field as a simple lattice integer field,
 and to reproduce the Kapustin--Seiberg prescription in the continuum limit.
 Our construction is also applied to analyzing the higher-group structure
 in the $SU(N)$ gauge theory with the instanton-sum modification.
\end{abstract}
\subjectindex{B01, B06, B31, B35}
\maketitle

\section{Introduction and brief review of coupling to higher-form gauge fields}
To study the dynamics of gauge fields has been a profound problem for a long time.
A traditional gauge principle arises from the localization process
of a global symmetry for phases (or elements of group) of matters~\cite{Pauli:1941zz};
this procedure is called gauging.
A global symmetry in quantum field theory may be gauged as a probe,
and then there is an 't~Hooft anomaly
when such a gauge symmetry is anomalous~\cite{tHooft:1979rat}.
The 't~Hooft anomalies in low and high energies
should be matched because of its invariance under the renormalization group flow,
which restricts the phase structure of strongly coupled gauge theories.

In the last decade, the concept of symmetry has been
generalized~\cite{Gaiotto:2014kfa,Gaiotto:2017yup}.
This so-called generalized global symmetry has been vigorously studied
in not only particle physics but also condensed matter
physics~\cite{Cordova:2022ruw,McGreevy:2022oyu,Gomes:2023ahz}.
The important ingredients are as follows:
\begin{itemize}
 \item higher-form symmetry: when a theory has symmetries acting
 on not only local operators but also a $p$-dimensional \emph{charged} object,
 such a symmetry is called the $p$-form symmetry;
 \item higher-group symmetry: a categorical structure between some
 higher-form symmetries is realized~\cite{Kapustin:2013qsa,Kapustin:2013uxa},
 where each symmetry cannot be gauged individually;\footnote{%
 E.g., for $0$-form and $1$-form symmetries,
 the $2$-form gauge field associated to the $1$-form symmetry
 transforms under not only the $1$-form gauge transformation
 but also the $0$-form gauge transformation.
 If those symmetries are continuous, this is nothing but
 the Green--Schwarz mechanism~\cite{Green:1984sg}.}
 \item non-invertible symmetry: a symmetry, which cannot be represented
 by a symmetry \emph{group}, is given by a fusion rule
 between topological defects~\cite{Verlinde:1988sn,Bhardwaj:2017xup,%
 Chang:2018iay,Koide:2021zxj,Choi:2021kmx,Kaidi:2021xfk}\footnote{%
 For recent works related to our approach in this paper,
 see also Refs.~\cite{Bhardwaj:2021wif,Bhardwaj:2022kot,%
 Bhardwaj:2022maz,Bhardwaj:2022yxj,Delcamp:2023kew}.}
 so that there exist no inverse topological operators;
\end{itemize}
etc.
The recent developments in line with these symmetries
provide a quite different paradigm.

It is well known that
the $SU(N)$ gauge theory has the $\mathbb{Z}_N$ $1$-form center symmetry.
In order for non-Abelian gauge theories to couple
with $\mathbb{Z}_N$ $2$-form gauge fields associated with such $1$-form symmetries,
we can use the following procedure by Kapustin and
Seiberg~\cite{Kapustin:2014gua,Gaiotto:2014kfa,Gaiotto:2017yup}:
Let a topological field theory be described by a $\mathbb{Z}_N$ $p$-form gauge field.
To represent this explicitly, introducing a $(p-1)$-form compact scalar
$B^{(p-1)}$ which satisfies $B^{(p-1)}\sim B^{(p-1)}+2\pi$,
a $U(1)$ $p$-form field $B^{(p)}$, and a Lagrange multiplier~$\chi$,
we write the Lagrangian\footnote{%
We can regard $\chi$ as the vacuum expectation value (VEV)
of a charge-$N$ Higgs field~$H$,
and the compact scalar as the phase of $H$.
If we take the $\chi\to\infty$ limit (or infinite Higgs VEV),
we have the same structure.}
\begin{align}
 \chi\wedge\left(N B^{(p)}-\dd{B^{(p-1)}}\right) .
\end{align}
The corresponding charged object behaves as
\begin{align}
 e^{i\int_{\text{$p$-cycle}} B^{(p)}}
 = e^{\frac{i}{N}\int_{\text{$p$-cycle}} \dd{B^{(p-1)}}} \in \mathbb{Z}_N .
\end{align}
Thus, the $\mathbb{Z}_N$ gauge field is written by the pair ($B^{(p)}$, $B^{(p-1)}$).
Note that the $SU(N)$ gauge field~$A$ is blind
to such a $\mathbb{Z}_N$ $2$-form gauge field ($B^{(2)}$, $B^{(1)}$)
because $A$ (also the field strength~$F$) is traceless.
Now, \emph{let us promote $A$ to a $U(N)$ gauge field $\mathcal{A}$}.
In the same way above, we constrain its $U(1)$ part as
\begin{align}
 \chi'\wedge\left(\Tr\mathcal{F}-\dd{B^{(1)}}\right) .
\end{align}
This indicates that $U(1)$ is broken to~$\mathbb{Z}_N$, that is,
$U(N)=\frac{SU(N)\times U(1)}{\mathbb{Z}_N}\to SU(N)$.
Further, by replacing $F\to\mathcal{F}-B^{(2)}$, the $SU(N)$ gauge field can couple
to the $\mathbb{Z}_N$ $2$-form gauge field;
we have the $SU(N)/\mathbb{Z}_N$ gauge theory.

This recent prescription based on higher-form symmetries
could be described equivalently by the $SU(N)/\mathbb{Z}_N$ gauge theory
where the $SU(N)$ gauge fields obey
the twisted boundary condition with the 't~Hooft flux~\cite{tHooft:1979rtg}.
It is then known that the topological charge becomes fractional;
in terms of the $2$-form gauge field~$B^{(2)}$, we see easily
\begin{align}
 \int \dd[4]{x}\Tr\left(\mathcal{F}-B^{(2)}\right)\wedge\left(\mathcal{F}-B^{(2)}\right)
 = \int \dd[4]{x}\Tr\mathcal{F}\wedge\mathcal{F}
 - N \int \dd[4]{x} B^{(2)}\wedge B^{(2)} ,
\end{align}
where the first term gives rise to an integer in the topological charge,
and the second term can provide a fractional part
because $\int_{\text{$2$-cycle}} B^{(2)}\in\frac{2\pi}{N}\mathbb{Z}$.
The proof of the fractionality under the twisted boundary condition
was done by van~Baal~\cite{vanBaal:1982ag},
where the author constructs the transition function
with the $\mathbb{Z}_N$-center-valued cocycle condition
and then identifies the topological classification
of the $SU(N)$ principal bundle structure.

L\"uscher proved~\cite{Luscher:1981zq} that
the $SU(N)$ principal bundle can be constructed from the $SU(N)$ lattice gauge theory
(see also Refs.~\cite{Phillips:1986qd,Phillips:1990kj}),
while the topological structure on the lattice is nontrivial
because the discretization of the spacetime breaks its continuity.
We can recover this ``continuity'' by L\"uscher's construction,
and then, classify the integer topological charges.
Recall that quantum field theory,
a physical system with an infinite number of degrees of freedom,
would be mathematically not well-defined as it stands,
and the lattice regularization is the most well-developed non-perturbative framework.
This work is quite awesome since it provides a solid foundation
on our understanding of topological classifications,
which enrich the nontrivial dynamics of gauge fields
(e.g., the index theorem~\cite{Atiyah:1963zz,Atiyah:1968mp} and so on).

Recently, in Refs.~\cite{Abe:2022nfq,Abe:2023ncy}, by generalizing this,
the non-simply-connected $U(1)/\mathbb{Z}_N$
or $SU(N)/\mathbb{Z}_N$ principal bundle has been constructed
in the lattice theory coupled with the $\mathbb{Z}_N$ $2$-form gauge field,
and the fractionality of the topological charge on the lattice is proved.
These studies have achieved the fully regularized framework
on the modern viewpoint of quantum field theories with generalized symmetries.

The proof in Ref.~\cite{Abe:2023ncy} is given in a sophisticated way
based on the principle of the locality,
$SU(N)$ gauge invariance, and $\mathbb{Z}_N$ $1$-form gauge invariance,
while this idea looks different from that for the $U(1)$ case in~Ref.~\cite{Abe:2022nfq}
and the construction is quite complicated.
Thus, from the analytical viewpoint in lattice gauge theory,
it may be hard to explicitly demonstrate
the traditional knowledge on the 't~Hooft twisted boundary condition, and
recent developments of non-Abelian gauge theories with higher-form symmetries.
Also the relation between the Kapustin--Seiberg prescription by the $U(1)$ fields
and the lattice construction by the $2$-form integer lattice field is not obvious;
it is puzzling how to take the continuum limit of an integer field on the lattice
so that its field configuration is smooth.

In this paper, we reconstruct the $SU(N)/\mathbb{Z}_N$ principal bundle
from the lattice $SU(N)$ gauge theory with the twisted boundary condition.
First, we start from the continuum theory following van~Baal~\cite{vanBaal:1982ag},
on which the construction of the $U(1)/\mathbb{Z}_N$ principal bundle~\cite{Abe:2022nfq} is based.
We can make his discussions much simpler owing to the extension
of $SU(N)$ to $U(N)$ like as the Kapustin--Seiberg prescription.
Actually, the original proof seems to be incompatible
with interpolated transition functions
written by lattice $SU(N)$ gauge fields,
but the above description is applicable to both continuum and lattice.
Next, we give the lattice realization of it,
by using the generalized L\"uscher's construction~\cite{Abe:2023ncy}.
We can then show how to establish an equivalence between the constructions in this paper and Refs.~\cite{Abe:2022nfq,Abe:2023ncy}.
The distinguishing feature is that our construction provides concrete expressions
for $\mathcal{F}$ and $B^{(2)}$ defined on the lattice
while those are not defined in Ref.~\cite{Abe:2023ncy}.
That is, this point is not necessary
to prove the fractionality of the topological charge,
but we can see the fractional structure as an explicit form
and apply this construction to some related issues.

As an important perspective from our construction,
we perform the $U(1)$ $1$-form gauge transformation,
to which the gauged $\mathbb{Z}_N$ $1$-form center symmetry is promoted.
We then directly reproduce the Kapustin--Seiberg prescription
in terms of lattice fields.
Also we apply our expressions to analyzing the higher-group structure
in the $SU(N)$ gauge theory with the so-called instanton-sum
modification~\cite{Tanizaki:2019rbk},
which is a restriction of the topological sectors
to the instanton numbers characterized by an integer without violating
the locality~\cite{Pantev:2005rh,Pantev:2005wj,Pantev:2005zs,Seiberg:2010qd}.
We expect that the procedure in this paper is applicable broadly
as an underlying fully regularized framework.

\section{A revision of the $SU(N)$ gauge theory with twisted boundary condition}
\label{sec:vanBaal}
At first, we review how the topological charge becomes fractional
in the $SU(N)$ gauge theory with a twisted boundary condition on a torus.
The original proof by van~Baal~\cite{vanBaal:1982ag} is modified in a simpler way,
in accordance with the recent viewpoint on non-Abelian gauge theories coupled
with $\mathbb{Z}_N$ gauge fields~\cite{Kapustin:2014gua}.

A four-dimensional periodic torus with the size~$L$ is give by
\begin{align}
 T^4&\equiv\left\{x\in\mathbb{R}^4 \mid \text{$0\leq x_\mu<L$ for all $\mu$}\right\}
\end{align}
with the identification~$x_\mu+L\sim x_\mu$,
where $\mu$ runs over $1$, $2$, $3$, and $4$.
We impose the twisted boundary condition
for the gauge field~$A_\mu(x)\in\mathfrak{su}(N)$,
\begin{align}
 A_\mu(x_\nu+L, x_{\lambda\neq\nu})
 = h_\nu(x)^{-1} A_\mu(x) h_\nu(x)
 - i h_\nu(x)^{-1} \partial_\mu h_\nu(x) ,
\end{align}
with the $\mathbb{Z}_N$-relaxed cocycle condition
\begin{align}
 &h_\mu(x_\mu=0,x_\nu=L,x_{\lambda\neq\mu,\nu})^{-1}
 h_\nu(x_\mu=x_\nu=0,x_{\lambda\neq\mu,\nu})^{-1}
 \notag\\&\qquad\times
 h_\mu(x_\mu=x_\nu=0,x_{\lambda\neq\mu,\nu})
 h_\nu(x_\mu=L,x_\nu=0,x_{\lambda\neq\mu,\nu})
 = \exp\left(\frac{2\pi i}{N}z_{\mu\nu}\right)
 \in\mathbb{Z}_N .
 \label{eq:vanBaal_twist_bc}
\end{align}
Here $z_{\mu\nu}\in\mathbb{Z}$, which is anti-symmetric as~$z_{\mu\nu}=-z_{\nu\mu}$,
stands for the 't~Hooft flux~\cite{tHooft:1979rtg}.

The topological charge is specified by the transition function~$h_\mu(x)$.
We see the Lemma proved by van~Baal
and L\"uscher~\cite{Luscher:1981zq,vanBaal:1982ag} as follows:
\begin{lem*}
Subject to the twisted boundary condition~\eqref{eq:vanBaal_twist_bc}
(or the periodic one with~$z_{\mu\nu}=0$),
the topological charge is written, in terms of~$h_\mu(x)$, by
\begin{align}
 Q &= \frac{1}{32\pi^2} \int_{T^4} \dd[4]{x}
 \sum_{\mu,\nu,\rho,\sigma}\epsilon_{\mu\nu\rho\sigma}
 \Tr\left[F_{\mu\nu}(x)F_{\rho\sigma}(x)\right]
 \notag\\
 &= - \frac{1}{24\pi^2} \sum_{\mu,\nu,\rho,\sigma}
 \epsilon_{\mu\nu\rho\sigma}
 \biggl\{
 3 \int \dd{x_\rho} \dd{x_\sigma}
 \Tr\left[\left(h_\mu\partial_\rho h_\mu^{-1}\right)_{x_\mu=x_\nu=0}
 \left(h_\nu^{-1}\partial_\sigma h_\nu\right)_{x_\mu=L,x_\nu=0}\right]
 \notag\\&\qquad\qquad\qquad\qquad\qquad
 + \int \dd{x_\nu} \dd{x_\rho} \dd{x_\sigma}\Tr\left[
 \left(h_\mu^{-1}\partial_\nu h_\mu\right)
 \left(h_\mu^{-1}\partial_\rho h_\mu\right)
 \left(h_\mu^{-1}\partial_\sigma h_\mu\right)
 \right]_{x_\mu=0}
 \biggr\} .
\end{align}
\end{lem*}

\begin{figure}
  \centering
  \begin{tikzpicture}[scale=1.2]
    \draw (0,0) rectangle (8,4);
    \foreach \x in {0,1}{
        \foreach \y in {0,1}{
            \node at (0.5+7*\x,0.5+3*\y) {$U_3$};
            \node at (0.5+7*\x,2) {$U_1$};
            \node at (4,0.5+3*\x) {$U_2$};
            \draw [dashed] (0,1+2*\x)--(8,1+2*\x)  (1+6*\x,0)--(1+6*\x,4);
            \draw [<-, red, very thick] (0.25+7*\x,0.5+3*\y)--(-0.25+7*\x,0.5+3*\y) node [left] {$h_{23}$};
            \draw [<-, blue, very thick] (0.75+7*\x,0.5+3*\y)--(1.25+7*\x,0.5+3*\y) node [right] {$g_{23}$};
            \draw [<-, blue, very thick] (0.5+7*\x,0.75+3*\y)--(0.5+7*\x,1.25+3*\y) node [above] {$g_{13}$};
            \draw [<-, red, very thick] (0.5+7*\x,0.25+3*\y)--(0.5+7*\x,-0.25+3*\y) node [below] {$h_{13}$};
            \draw [<-, blue, very thick] (4,0.75+3*\y)--(4,1.25+3*\y) node [above] {$g_{02}$};
            \draw [<-, red, very thick] (4,0.25+3*\y)--(4,-0.25+3*\y) node [below] {$h_{02}$};
            \draw [<-, blue, very thick] (0.75+7*\x,2)--(1.25+7*\x,2) node [right] {$g_{01}$};
            \draw [<-, red, very thick] (0.25+7*\x,2)--(-0.25+7*\x,2) node [left] {$h_{01}$};
            \fill [radius=2pt] (1+6*\x,1+2*\y) circle;
        }
    };
    \node at (4,2) {$U_0$};
    \draw [->,>=stealth] (-3,0) -- (-2,0) node [right] {$x_1$};
    \draw [->,>=stealth] (-3,0) -- (-3,1) node [above] {$x_2$};
    \draw [<->,>=stealth] (0,-1) -- (1,-1) node [right] {$\delta$};
    \draw [<->,>=stealth] (-1.15,0) -- (-1.15,1) node [above] {$\epsilon$};
  \end{tikzpicture}
  \caption{The structure of the fiber bundle of the base space $T^2$.
  $\{U_i\}$ is a set of patches of~$T^2$, where $U_0$ is the main region,
  $U_\mu$ is near the boundary at $x_\mu=L$,
  and $U_3$ is the corner.}
  \label{fig:vanBaal_bundle}
\end{figure}
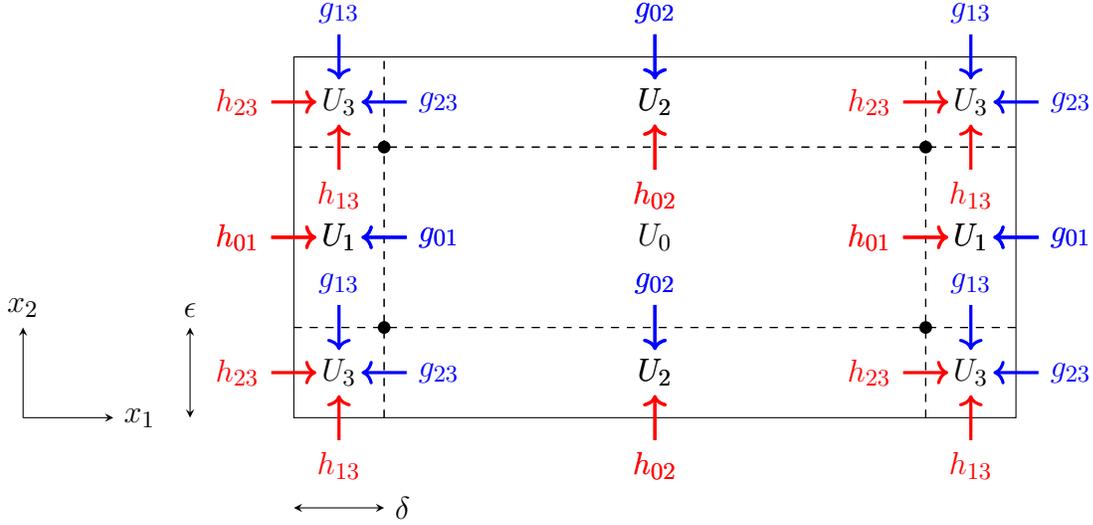

If we have no twists ($z_{\mu\nu}=0$), $Q\in\mathbb{Z}$ characterizes
the homotopy type $\pi_3(SU(N))=\mathbb{Z}$.
On the other hand, turning on $z_{\mu\nu}$,
we can see that $Q$ becomes fractional
by considering an appropriate fiber bundle structure
due to the first homotopy group $\pi_1(SU(N)/\mathbb{Z}_N)=\mathbb{Z}_N$ on~$T^4$.
To see this, for simplicity,
we consider the bundle structure of $T^2$ as depicted in~Fig.~\ref{fig:vanBaal_bundle}.
Letting $U_i$ with $\epsilon>0$ and $\delta>0$ be a covering of~$T^2$,
$h_{ij}$ and $g_{ij}$ are transition functions at $x\in U_i\cap U_j$
such that $A_\mu(U_i)\to A_\mu(U_j)$
in the positive and negative $x$ directions, respectively.
In the limit of $\epsilon\to0$ and $\delta\to0$,
the patches $U_{i\neq0}$ shrink
and the nontrivial transition indicates that
$A_\mu(U_0)\to A_\mu(U_\nu)\to A_\mu(U_0)$,
that is, $A_\mu(x_\nu+L,x_{\lambda\neq\nu})\mapsto A_\mu(x)$.
Therefore, on~$T^4$, we have $h_\mu=h_{0\mu}g_{0\mu}^{-1}$
with an appropriate assignment of indices.

Now, following the van~Baal's prescription~\cite{vanBaal:1982ag},
we introduce the loop factor~$\Tilde\varsigma_\mu(x)$
in terms of the Cartan subalgebra of $SU(N)$,
\begin{align}
 \Tilde\varsigma_\mu(x) \equiv
 \exp\left(-\frac{2\pi i}{N}\sum_{\nu>\mu}\frac{z_{\mu\nu}x_\nu}{L}T_1\right) ,
\end{align}
where $T_1$ is a generator of~$SU(N)$,
\begin{align}
 T_1 \equiv \mathrm{diag}(1,1,\dots,1,-N+1) .
\end{align}
Then we find
\begin{align}
 &\Tilde\varsigma_\mu(x_\mu=0,x_\nu=L,x_{\lambda\neq\mu,\nu})^{-1}
 \Tilde\varsigma_\nu(x_\mu=x_\nu=0,x_{\lambda\neq\mu,\nu})^{-1}
 \notag\\&\qquad\times
 \Tilde\varsigma_\mu(x_\mu=x_\nu=0,x_{\lambda\neq\mu,\nu})
 \Tilde\varsigma_\nu(x_\mu=L,x_\nu=0,x_{\lambda\neq\mu,\nu})
 = \exp\left(\frac{2\pi i}{N}z_{\mu\nu}\right)
 \in\mathbb{Z}_N .
 \label{eq:cocycle_sigma}
\end{align}
This cocycle condition is identical
to that of~$h_\mu(x)$ \eqref{eq:vanBaal_twist_bc},
and thus, $\Tilde\varsigma_\mu(x)$ possesses
the same nontrivial ``winding'' modulo $1$ as the original system.

\begin{thm*}
Given $\forall h_\mu(x)$, we can write
\begin{align}
 h_\mu(x) = h_{0\mu}g_{0\mu}^{-1}
 = \Hat{h}_{0\mu}\Tilde\varsigma_\mu\Hat{g}_{0\mu}^{-1} ,
\end{align}
where we have introduced
\begin{align}
 \Tilde\varsigma_\mu &= \Tilde{h}_{0\mu}\Tilde{g}_{0\mu}^{-1} ,&
 \Hat{h}_{0\mu} &= h_{0\mu}\Tilde{h}_{0\mu}^{-1} ,&
 \Hat{g}_{0\mu} &= g_{0\mu}\Tilde{g}_{0\mu}^{-1} ,
\end{align}
and $\Hat{h}_{0\mu}$, $\Hat{g}_{0\mu}$ are defined
in terms of the transition function
$\Hat{h}_\mu(x)=\Hat{h}_{0\mu}\Hat{g}_{0\mu}^{-1}$
obeying the periodic boundary condition.
\end{thm*}
\begin{cor*}
Let $Q[\cdot]$ be a mapping from a transition function
to the topological charge given by the above Lemma.
Then, we have
\begin{align}
 Q[h] = Q[\Tilde\varsigma] + Q[\Hat{h}] .
\end{align}
\end{cor*}

The proof being given in Ref.~\cite{vanBaal:1982ag},
we can then show that $Q[\Tilde\varsigma]\in\frac{1}{N}\mathbb{Z}$
and $Q[\Hat{h}]\in\mathbb{Z}$.
Thus the total topological charge
on the $SU(N)/\mathbb{Z}_N$ principal bundle can be fractional.

The above original construction is quite complicated
because its description enjoys the $SU(N)$ structure at any stage of computations.
Also, lattice gauge theory as we will describe later
seems to be incompatible with the above Theorem.
If such a difficulty of $SU(N)$ keeps us in mind of the recent development about
the generalized symmetry~\cite{Kapustin:2014gua},
one may begin with the loop factor $\varsigma_\mu(x)\in U(1)$ redefined by
\begin{align}
 \varsigma_\mu(x) \equiv
 \exp\left(-\frac{2\pi i}{N}\sum_{\nu>\mu}\frac{z_{\mu\nu}x_\nu}{L}\right) ,
\end{align}
and the $U(N)$ structure which is combined with~$\varsigma_\mu(x)$
into $SU(N)$-valued results.
This definition of $\varsigma_\mu(x)$ again satisfies
the cocycle condition~\eqref{eq:cocycle_sigma}.
By using $\varsigma_\mu(x)\in U(1)$, let us define
\begin{align}
 \Check{h}_\mu(x) \equiv \varsigma_\mu(x)^{-1} h_\mu(x) \in U(N).
\end{align}
Then, we immediately find that
\begin{align}
 &\Check{h}_\mu(x_\mu=0,x_\nu=L,x_{\lambda\neq\mu,\nu})^{-1}
 \Check{h}_\nu(x_\mu=x_\nu=0,x_{\lambda\neq\mu,\nu})^{-1}
 \notag\\&\qquad\times
 \Check{h}_\mu(x_\mu=x_\nu=0,x_{\lambda\neq\mu,\nu})
 \Check{h}_\nu(x_\mu=L,x_\nu=0,x_{\lambda\neq\mu,\nu})
 = 1 .
\end{align}
$\Check{h}_\mu(x)$ denotes the transition function
of the ``$U(N)$ principal bundle'' with an integer 2nd Chern number.

Substituting $h_\mu(x)=\varsigma_\mu(x)\Check{h}_\mu(x)$
to the expression of the topological charge,
and using the integer topological charge $\Check{Q}$
by~$\Check{h}_\mu(x)$ instead of $h_\mu(x)$ in the above expression of~$Q$,\footnote{%
This integer topological charge for the $U(N)$ gauge field, $\int\Tr F\wedge F$,
is somewhat different from the definition of the 2nd Chern class,
$\int(\Tr F\wedge F-\Tr F\wedge\Tr F)$, up to integers.}
we have
\begin{align}
 Q &= \frac{1}{8N} \sum_{\mu,\nu,\rho,\sigma}
 \epsilon_{\mu\nu\rho\sigma} z_{\mu\nu}z_{\rho\sigma}
\notag\\&\qquad
 - \frac{i}{4\pi N} \sum_{\mu,\nu,\rho>\mu,\sigma}
 \epsilon_{\mu\nu\rho\sigma} \frac{z_{\mu\rho}}{L}
 \int \dd{x_\rho} \dd{x_\sigma}
 \Tr\left[
 \left(\Check{h}_\nu^{-1}\partial_\sigma\Check{h}_\nu\right)_{x_\mu=L,x_\nu=0}
 - \left(\Check{h}_\nu\partial_\sigma\Check{h}_\nu^{-1}\right)_{x_\mu=x_\nu=0}
 \right]
\notag\\&\qquad
 + \Check{Q}
\notag\\
 &= - \frac{1}{8N} \sum_{\mu,\nu,\rho,\sigma}
 \epsilon_{\mu\nu\rho\sigma} z_{\mu\nu}z_{\rho\sigma}
 + \Check{Q} .
\end{align}
In the last line, we have used, for $\mu<\nu$,
\begin{align}
 0 = \Tr\left(h_\mu^{-1}\partial_\nu h_\mu\right)
 = - 2\pi i\frac{z_{\mu\nu}}{L}
 + \Tr\left(\Check{h}_\mu^{-1}\partial_\nu\Check{h}_\mu\right) .
 \label{eq:crossterm}
\end{align}
The first term in~$Q$ is fractional, that is,
it is proved that $Q\in\frac{1}{N}\mathbb{Z}$;
the fractional part and $\Check{Q}\in\mathbb{Z}$ are naturally decomposed.

\section{Lattice $SU(N)$ gauge theory with twisted boundary condition}
\subsection{Setup as a lattice formulation}
It would be difficult to consider the topological structure
in lattice gauge theories, where the spacetime is discretized
into a set of lattice points and then its continuity is broken.
We can recover this feature by the L\"uscher's construction
of the $SU(N)$ principal bundle
from the lattice $SU(N)$ gauge theory~\cite{Luscher:1981zq}.
Quite recently, this construction has been generalized
to non-Abelian gauge theories coupled
with $\mathbb{Z}_N$ gauge fields~\cite{Abe:2023ncy}.
Let us demonstrate the $SU(N)/\mathbb{Z}_N$ principal bundle
in line with the van~Baal's proof in the previous section.

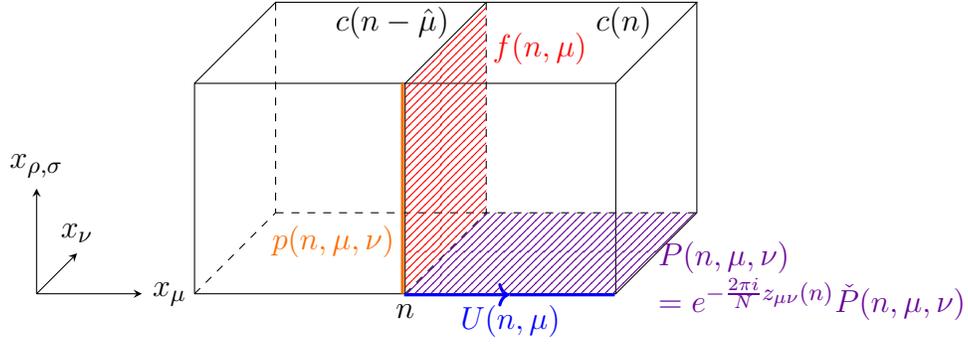
\begin{figure}
    \centering
    \begin{tikzpicture}[scale=0.7]
        \draw (0,0,0)--(0,4,0)--(4,4,0)--(4,0,0)--(0,0,0);
        \draw (4,4,0)--(4,4,-4)--(0,4,-4)--(0,4,0);
        \draw (4,0,0)--(8,0,0)--(8,4,0);
        \draw (4,4,0)--(8,4,0)--(8,4,-4)--(4,4,-4);
        \draw (8,0,0)--(8,0,-4)--(8,4,-4);
        \draw[dashed](0,0,0)--(0,0,-4)--(0,4,-4);
        \draw[dashed](0,0,-4)--(4,0,-4)--(4,0,0);
        \draw[dashed](4,4,-4)--(4,0,-4)--(8,0,-4);
        \draw [red!15!orange, very thick] (3.95,0,0)--(3.95,4,0);
        \draw [->,blue, very thick] (4,0,0.05)--(6,0,0.05);
        \draw [blue, very thick] (6,0,0.05)--(8,0,0.05);
        \fill[pattern=north east lines, pattern color = red](4,0,0)--(4,0,-4)--(4,4,-4)--(4,4,0);
        \node [red] at (5.4,3.5,-3) {$f(n,\mu)$};
        \fill[pattern=north east lines, pattern color = red!40!blue](4,0,0)--(8,0,0)--(8,0,-4)--(4,0,-4);
        \node [red!40!blue] at (8.6,0.7,0) [right] {$P(n,\mu,\nu)$};
        \node [red!40!blue] at (8.6,-0.1,0) [right] {$=e^{-\frac{2\pi i}{N}z_{\mu\nu}(n)}\Check{P}(n,\mu,\nu)$};
        \node at (4,0,0) [below] {$n$};
        \node [red!15!orange] at (4,1,0) [left] {$p(n,\mu,\nu)$};
        \node [blue] at (6,0,0) [below] {$U(n,\mu)$};
        \node at  (2.6,4,-3) {$c(n-\Hat{\mu})$};
        \node at (7,4,-3) {$c(n)$};
        \draw[->,>=stealth] (-3,0,0) -- (-1,0,0) node [right]{$x_\mu$};
        \draw[->,>=stealth] (-3,0,0) -- (-3,0,-2) node [above]{$x_\nu$};
        \draw[->,>=stealth] (-3,0,0) -- (-3,2,0) node [above]{$x_{\rho,\sigma}$};
    \end{tikzpicture}
    \caption{Lattice setup and plaquettes for the twisted and periodic link variables.
    To illustrate generic variables, we use a lattice field
    $z_{\mu\nu}(n)=z_{\mu\nu}\delta_{n_\mu,L-1}\delta_{n_\nu,L-1}$
    as we will define later.}
    \label{fig:lat_setup}
\end{figure}

The lattice~$\Lambda$,
\begin{align}
 \Lambda&\equiv\left\{n\in\mathbb{Z}^4
 \mid \text{$0\leq n_\mu<L$ for all $\mu$}\right\} ,
\end{align}
divides $T^4$ into hypercubes~$c(n)$ as
\begin{align}
 c(n)\equiv\left\{x\in\mathbb{R}^4
 \mid \text{$0\leq(x_\mu-n_\mu)\leq1$ for all $\mu$}\right\} .
\end{align}
We also define the boundary of two hypercubes, called the face,
\begin{align}
 f(n,\mu)\equiv\left\{x\in c(n) \mid x_\mu=n_\mu\right\} = c(n-\Hat{\mu})\cap c(n),
\end{align}
where $\Hat{\mu}$ is a unit vector in the positive $\mu$ direction,
and a two-dimensional plaquette as the intersection of four hypercubes,
$c(n)$, $c(n-\Hat{\mu})$, $c(n-\Hat{\nu})$, and~$c(n-\Hat{\mu}-\Hat{\nu})$:
\begin{align}
 p(n,\mu,\nu)\equiv\left\{x\in c(n) \mid \text{$x_\mu=n_\mu$, $x_\nu=n_\nu$}\right\}
 \qquad (\mu\neq\nu).
\end{align}
Our lattice setup is summarized in~Fig.~\ref{fig:lat_setup}.

Suppose that the link variable~$U(n,\mu)\in SU(N)$,
which lives on the link connecting $n$ and~$n+\Hat{\mu}$,
obeys the twisted boundary condition,
\begin{align}
 U(n+L\Hat\nu,\mu)=g_\nu(n)^{-1}U(n,\mu)g_\nu(n+\Hat\mu).
\end{align}
The cocycle condition is given by
\begin{align}
 g_\mu(n+L\Hat\nu)^{-1}g_\nu(n)^{-1}g_\mu(n)g_\nu(n+L\Hat\mu)
 = \exp\left({\frac{2\pi i}{N}z_{\mu\nu}}\right) \in \mathbb{Z}_N.
\end{align}
This represents the 't~Hooft flux on the lattice.

To rewrite the link variable $U(n,\mu)$ in terms of the periodic one,
we assume that $U(n,\mu)$ is defined on~$0\leq n_{\nu\neq\mu}\leq L$
and~$0\leq n_\mu\leq L-1$.
We then define the periodic variable~$\Check{U}(n,\mu)$ by
\begin{align}
 U(n,\mu)
 =
 \begin{cases}
  \Check{U}(n,\mu) g_\mu(n) & \text{for $n_\mu=L-1$},\\
  \Check{U}(n,\mu) & \text{otherwise}.
 \end{cases}
\end{align}
One can find that
\begin{align}
 & g_\mu(n_\nu=L-1)
 U(n_\mu=L,n_\nu=L-1,\nu) U(n_\mu=L-1,n_\nu=L,\mu)^{-1}
 g_\nu(n_\mu=L-1)^{-1}
 \notag\\
 &= \exp\left(-\frac{2\pi i}{N}z_{\mu\nu}\right)
 \Check{U}(n_\mu=0,n_\nu=L-1,\nu)\Check{U}(n_\mu=L-1,n_\nu=0,\mu)^{-1} .
\end{align}
This shows that, as depicted in Fig.~\ref{fig:lat_setup},
the plaquette~$P(n,\mu,\nu)$,
\begin{align}
 P(n,\mu,\nu)\equiv U(n,\mu)U(n+\Hat\mu,\nu)U(n+\Hat\nu,\mu)^{-1}U(n,\nu)^{-1},
\end{align}
can be written by
$P(n,\mu,\nu)=e^{-\frac{2\pi i}{N}z_{\mu\nu}}\Check{P}(n,\mu,\nu)$
up to gauge functions
if this Wilson loop passes the corner of the lattice,
where $\Check{P}(n,\mu,\nu)$ is the plaquette constructed by~$\Check{U}(n,\mu)$;
otherwise, $P(n,\mu,\nu)=\Check{P}(n,\mu,\nu)$.
In what follows, we simply say
$P(n,\mu,\nu)=e^{-\frac{2\pi i}{N}z_{\mu\nu}}\Check{P}(n,\mu,\nu)$
as this meaning.

\subsection{$SU(N)/\mathbb{Z}_N$ principal bundle and fractionality on the lattice}\label{sec:3.2}
Let us apply the L\"uscher's construction of the transition function
to~$U(n,\mu)$ and~$\Check{U}(n,\mu)$,
whose transition functions are denoted
by $v_{n,\mu}(x)$ and $\Check{v}_{n,\mu}(x)$, respectively.
We aim to write $v_{n,\mu}(x)$ in terms of~$\Check{v}_{n,\mu}(x)$.
A transition function $g_{n,\mu}(u; x)$ for a link variable~$u(n,\mu)$
at~$\forall x\in f(n,\mu)$ is defined by,
in terms of the L\"uscher's interpolation function
$S_{n,\mu}^m(u; x)$~\cite{Luscher:1981zq}
(for explicit formulas, see Appendix~\ref{sec:luscher_formulas}),
\begin{align}
 g_{n,\mu}(u; x)
 &\equiv S_{n,\mu}^{n-\hat\mu}(u; x)^{-1} g_{n,\mu}(u; n) S_{n,\mu}^n(u; x)
 \notag\\
 &= S_{n,\mu}^{n-\hat\mu}(u; x)^{-1} w^{n-\hat\mu}(u; n)
 w^n(u; n)^{-1} S_{n,\mu}^n(u; x) ,
\end{align}
where the standard parallel transporter $w^n(u; x)$
at the corners of~$f(n,\mu)$ is given by
\begin{align}
 w^n(u; x) = u(n,4)^{y_4} u(n+y_4\Hat{4},3)^{y_3}
 u(n+y_4\Hat{4}+y_3\Hat{3},2)^{y_2}
 u(n+y_4\Hat{4}+y_3\Hat{3}+y_2\Hat{2},1)^{y_1}
 \notag\\
 \text{for $y_\mu\equiv x_\mu-n_\mu=0$ or $1$}.
 \label{eq:parallel_trans}
\end{align}
Then, we naively define $v_{n,\mu}(x)=g_{n,\mu}(U; x)$
and $\Check{v}_{n,\mu}(x)=g_{n,\mu}(\Check{U}; x)$.
$v_{n,\mu}(x)$ obeys the twisted boundary condition,
\begin{align}
 v_{n,\mu}(x+L\Hat\nu) = g_\nu(n-\Hat\mu)^{-1} v_{n,\mu}(x) g_\mu(n) ,
 \label{eq:twist_bc}
\end{align}
while $\Check{v}_{n,\mu}(x)$ is periodic.

To obtain the well-defined interpolation $S_{n,\mu}^m(U;x)$,
we should impose an \emph{admissibility} condition.
Here, for simplicity, we consider the $SU(2)$ gauge theory
(we can generalize discussions below to any compact gauge group).\footnote{%
Admissible configurations should be close to those at the minimum in the classical continuum limit.
Thus, such configurations are topologically on a disk.
Since $SU(2)$ is topologically a sphere, removing simply one point on it,
we have a desired admissibility.}
First note that, to make the lattice action density small,
the plaquette $P$ is in a neighborhood of~$1$,
and $\Check{P}$ is $e^{\frac{2\pi i}{N}z_{\mu\nu}}$ or~$1$.
$S_{n,\mu}^m(U;x)$ is a function with respect to the plaquettes $P$,
and has a structure such as $P^y$ with $0\leq y\leq1$
(see Appendix~\ref{sec:luscher_formulas}).
Then, $P=-1$ is ill-defined and so such configurations are called exceptional.
Supposing all combinations in~$S_{n,\mu}^m(U;x)$ are well-defined,
we have the admissibility condition
\begin{align}
 \Tr\left[1 - P\right] < \varepsilon .
\end{align}
In Ref.~\cite{Abe:2023ncy},
it is proved that there exists $\varepsilon>0$ for~$\forall N$.
On the other hand, for $S_{n,\mu}^m(\Check{U};x)$,
the situation is more complicated because we cannot easily choose
a branch of $(\Check{P})^y\sim(e^{\frac{2\pi i}{N}z_{\mu\nu}})^y$.
To this end, let us construct $S_{n,\mu}^m(\Check{U};x)$ from~$S_{n,\mu}^m(U;x)$.
That is, since $P$ can be rewritten as $e^{-\frac{2\pi i}{N}z_{\mu\nu}}\Check{P}$,
we can define
\begin{align}
 \left(e^{-\frac{2\pi i}{N}z_{\mu\nu}}\Check{P}\right)^y
 = e^{-\frac{2\pi i}{N}(z_{\mu\nu}+N M_{\mu\nu})y}
 \left(e^{2\pi i M_{\mu\nu}}\Check{P}\right)^y ,
 \label{eq:branch}
\end{align}
where
\begin{align}
 \begin{cases}
  0\leq z_{\mu\nu}+N M_{\mu\nu}<N & \text{for $\mu<\nu$} ,\\
  z_{\mu\nu}+N M_{\mu\nu}=-z_{\nu\mu}-N M_{\nu\mu} & \text{for $\mu>\nu$} .
 \end{cases}
 \label{eq:z_range}
\end{align}
Also for a product of $k$ plaquettes,
\begin{align}
 \left(\prod_{\ell=1}^k e^{-\frac{2\pi i}{N}z_\ell}\Check{P}_\ell\right)^y
 = e^{-\frac{2\pi i}{N}\sum_{\ell=1}^k(z_\ell+N M_\ell)y}
 \left(e^{2\pi i\sum_{\ell=1}^k M_\ell}\prod_{\ell=1}^k\Check{P}_\ell\right)^y ,
 \label{eq:branch_k}
\end{align}
where $z_\ell=z_{\mu_\ell\nu_\ell}$ with Eq.~\eqref{eq:z_range}.
Note that $|\sum_{\ell}(z_\ell+N M_\ell)|<N k$.
In what follows, we redefine $z_{\mu\nu}+N M_{\mu\nu}$ as $z_{\mu\nu}$
so that~$0\leq z_{\mu\nu}<N$ for $\mu<\nu$.
Again, $\Check{P}=e^{\frac{2\pi i}{N}z_{\mu\nu}}\times(-1)$ is not defined,
and thus we have the same admissibility condition given by
\begin{align}
 \Tr \left[1-e^{-\frac{2\pi i}{N}z_{\mu\nu}}\Check{P}\right]
 < \varepsilon .
 \label{eq:admissibility}
\end{align}

We should mention that $\Check{v}_{n,\mu}(x)$ is an element of~$U(N)$.
This is because $(\Check{P})^y\sim e^{\frac{2\pi i}{N}z_{\mu\nu}y}\in U(1)$.
On lattice sites, that is, at $y=0$ or $1$,
this factor becomes $\mathbb{Z}_N$ so $\Check{v}_{n,\mu}\in SU(N)$.
When we write $v_{n,\mu}(x)$ in terms of $\Check{v}_{n,\mu}(x)$,
the extra factor of~$z_{\mu\nu}$ in $v_{n,\mu}(x)$ appears from
Eqs.~\eqref{eq:branch} and \eqref{eq:branch_k},
which is similar to $\varsigma_\mu(x)$ in the continuum theory.
This factor, $e^{-\frac{2\pi i}{N}z_{\mu\nu}y}$, is also an element of~$U(1)$.
These $U(1)$ factors, $(\Check{P})^y$ and $e^{-\frac{2\pi i}{N}z_{\mu\nu}y}$, cancel out by construction,
that is, $1\sim e^{-\frac{2\pi i}{N}z_{\mu\nu}y}(\Check{P})^y=P^y\in SU(N)$;
so $v_{n,\mu}(x)\in SU(N)$ is kept intact.

Following the above construction,
we can rewrite the transition function~$v_{n,\mu}(x)$
in terms of a $z_{\mu\nu}$ dependent factor and~$\Check{v}_{n,\mu}(x)$.
We obtain the transition function by, for~$x\in f(n,\mu)$,
\begin{align}
 v_{n,\mu}(x) =
 \begin{cases}
  \omega_{n,\mu}(x) \Check{v}_{n,\mu}(x) g_\mu(n-\Hat\mu)
  &\text{for $n_\mu=L$},\\
  \omega_{n,\mu}(x) \Check{v}_{n,\mu}(x) & \text{otherwise},
 \end{cases}
\end{align}
where the loop factor~$\omega_{n,\mu}(x)$ is defined by
\begin{align}
 \omega_{n,\mu}(x) \equiv
\begin{cases}
 \exp\left(\frac{2\pi i}{N}
 \sum_{\nu>\mu} z_{\mu\nu} y_\nu \delta_{n_\nu,L-1}\right)
 &\text{for $x_\mu=0\bmod L$} ,
 \\
 1 &\text{otherwise}.
\end{cases}
\end{align}
The transition functions provide the cocycle condition given by,
at~$x\in p(n,\mu,\nu)$,
\begin{align}
 v_{n-\Hat\nu,\mu}(x) v_{n,\nu}(x) v_{n,\mu}(x)^{-1} v_{n-\Hat\mu,\nu}(x)^{-1}
 &= 1 ,
\\
 \Check{v}_{n-\Hat\nu,\mu}(x)\Check{v}_{n,\nu}(x)
 \Check{v}_{n,\mu}(x)^{-1}\Check{v}_{n-\Hat\mu,\nu}(x)^{-1}
 &= 1 ,
 \label{eq:cocycle_check}
\\
 \omega_{n-\Hat\nu,\mu}(x) \omega_{n,\nu}(x)
 \omega_{n,\mu}(x)^{-1} \omega_{n-\Hat\mu,\nu}(x)^{-1}
 &=
 \begin{cases}
  \exp\left(\frac{2\pi i}{N} z_{\mu\nu}\right) & \text{$x_\mu=x_\nu=0\bmod L$,}
  \\
  1 & \text{otherwise}.
 \end{cases}
\end{align}
Note that $v_{n,\mu}(x)$ is nontrivial
because of the $\mathbb{Z}_N$ twisted boundary condition of it \eqref{eq:twist_bc}
though its cocycle condition is not relaxed by~$\mathbb{Z}_N$.
If one use $\Tilde{v}_{n,\mu}(x)=\omega_{n,\mu}(x)\Check{v}_{n,\mu}(x)$
obeying the periodic boundary condition, we find
\begin{align}
 \Tilde{v}_{n-\Hat\nu,\mu}(x) \Tilde{v}_{n,\nu}(x)
 \Tilde{v}_{n,\mu}(x)^{-1} \Tilde{v}_{n-\Hat\mu,\nu}(x)^{-1}
 &=
 \begin{cases}
  \exp\left(\frac{2\pi i}{N} z_{\mu\nu}\right) & \text{$x_\mu=x_\nu=0\bmod L$,}
  \\
  1 & \text{otherwise}.
 \end{cases}
\end{align}

Substituting $v_{n,\mu}(x)$ into the topological charge~$\mathcal{Q}$
\cite{vanBaal:1982ag,Luscher:1981zq}
\begin{align}
 \mathcal{Q} &= - \frac{1}{24\pi^2} \sum_{n\in\Lambda} \sum_{\mu,\nu,\rho,\sigma}
 \epsilon_{\mu\nu\rho\sigma}
 \biggl\{
 3 \int_{p(n,\mu,\nu)} \dd[2]{x}
 \Tr\left[\left(v_{n,\mu}\partial_\rho v_{n,\mu}^{-1}\right)
 \left(v_{n-\Hat\mu,\nu}^{-1}\partial_\sigma v_{n-\Hat\mu,\nu}\right)\right]
 \notag\\&\qquad\qquad\qquad\qquad\qquad\qquad
 + \int_{f(n,\mu)} \dd[3]{x} \Tr\left[
 \left(v_{n,\mu}^{-1}\partial_\nu v_{n,\mu}\right)
 \left(v_{n,\mu}^{-1}\partial_\rho v_{n,\mu}\right)
 \left(v_{n,\mu}^{-1}\partial_\sigma v_{n,\mu}\right)
 \right]
 \biggr\} ,
 \label{eq:topo_charge_lat}
\end{align}
we have
\begin{align}
 \mathcal{Q} = - \frac{1}{8N} \sum_{\mu,\nu,\rho,\sigma}
 \epsilon_{\mu\nu\rho\sigma}z_{\mu\nu}z_{\rho\sigma}
 + \Check{\mathcal{Q}} ,
 \label{eq:frac_topo}
\end{align}
where $\Check{\mathcal{Q}}$ is the topological charge with respect
to the periodic variable~$\Check{U}$, and we have used the similar identity
to Eq.~\eqref{eq:crossterm} for the cross terms.
The first term is a fractional part with~$\frac{1}{N}\mathbb{Z}$,
and one finds $\Check{Q}\in\mathbb{Z}$
thanks to the cocycle condition~\eqref{eq:cocycle_check}.

\subsection{Remarks on the relation with other constructions}
We make remarks on other constructions based on different ideas from ours:
$v_{n,\mu}(x)$ is similar to the transition function $\Tilde{v}_{n,\mu}(x)$ defined
in Ref.~\cite{Abe:2023ncy}
when we consider a specific 't~Hooft flux at the corner.
The different point is that $\Tilde{v}_{n,\mu}(x)$ is periodic.
Thus, $v_{n,\mu}(x) = \Tilde{v}_{n,\mu}(x) g_{\mu}(n-\Hat\mu)$ for~$n_\mu=L$.

Our definition of $\Check{v}_{n,\mu}(x)$ cannot be realized
as $g_{n,\mu}(\Check{U}; x)$ any longer without any other information.
Actually, it is not necessary to be based
on the underlying periodic variable~$\Check{U}(n,\mu)$.
As an alternative way, we can define $\Check{v}_{n,\mu}(x)$ as
$v_{n,\mu}(x)$ divided by the loop factor at the minimum,
as defined in the continuum by~$\varsigma_\mu(x)$ in Sect.~\ref{sec:vanBaal}.
The discussions in the paper can be also considered in this sense.

We ask how the definition of~$\Check{v}_{n,\mu}(x)$ is related to that
defined for the $U(1)$ lattice gauge theory in Ref.~\cite{Abe:2022nfq}.
The idea is, as an other prescription,
to replace the plaquette by it to the $N$th power;
that is, $\Check{P}^y\to(P^N)^{y/N}$
to impose the $\mathbb{Z}_N$ $1$-form gauge symmetry as we will define later.
This simple prescription removes the branch ambiguity above,
but is quite subtle and in fact not rigorous.
We first recognize that,
because $(e^{-\frac{2\pi i}{N}z_{\mu\nu}}\Check{P})^N=\Check{P}^N$,
we lose any information on the $\mathbb{Z}_N$ center,
and then $\Check{v}_{n,\mu}(x)=v_{n,\mu}(x)$
with the ``stronger'' admissibility~$\varepsilon/N$.
On the contrary, it looks like $e^{-\frac{2\pi i}{N}z_{\mu\nu}y}(\Check{P}^N)^{y/N}$,
and then, the information on the branch would be guessed
from the first factor thanks to the $\mathbb{Z}_N$ $1$-form gauge invariance.
Therefore, though this prescription is not a sure way,
it could hold up on a robust principle of the $1$-form gauge symmetry.

\section{$1$-form gauge invariance on the lattice}
\subsection{Gauging the $\mathbb{Z}_N$ $1$-form symmetry on the lattice}
We have started with the similar procedure to that in the continuum by van~Baal,
and obtained the fractional topological charge
with the 't~Hooft flux at the corner of~$\Lambda$.
Actually, it is known that
the structure of the $SU(N)/\mathbb{Z}_N$ principal bundle
from lattice theories is much more general and robust,
whose most important principle is based on the locality,
$SU(N)$ gauge invariance, and $\mathbb{Z}_N$ $1$-form gauge invariance.
In this section, let us consider the $\mathbb{Z}_N$ $1$-form
gauge invariance~\cite{Gaiotto:2014kfa,Kapustin:2013qsa,Kapustin:2014gua},
which plays a crucial role in this robustness~\cite{Abe:2023ncy}.

From the constant flux~$z_{\mu\nu}$, we introduce a lattice field $z_{\mu\nu}(n)$,
as a $\mathbb{Z}_N$ $2$-form gauge field,
\begin{align}
 z_{\mu\nu}(n)\equiv z_{\mu\nu}\delta_{n_\mu,L-1}\delta_{n_\nu,L-1} .
\end{align}
The loop factor is then given by
\begin{align}
 \omega_{n,\mu}(x) = \exp\left[
 \frac{2\pi i}{N}\sum_{\nu>\mu}z_{\mu\nu}(n-\Hat\mu)y_\nu\right],
\end{align}
and the twisted cocycle condition is
\begin{align}
 v_{n-\Hat\nu,\mu}(x)v_{n,\nu}(x)v_{n,\mu}(x)^{-1}v_{n-\Hat\mu,\nu}(x)^{-1}
 = \exp\left[\frac{2\pi i}{N}z_{\mu\nu}(n-\Hat\mu-\Hat\nu)\right] .
 \label{eq:cocycle_mod}
\end{align}

We give the $\mathbb{Z}_N$ $1$-form gauge transformation:
\begin{align}
 \Check{U}(n,\mu) \mapsto \exp\left[\frac{2\pi i}{N}z_\mu(n)\right] \Check{U}(n,\mu)
 \qquad z_\mu(n)\in\mathbb{Z},\, 0\leq z_\mu(n)<N.
\end{align}
Moreover, we assume
\begin{align}
 z_{\mu\nu}(n) \mapsto z_{\mu\nu}(n)
 + \Delta_\mu z_\nu(n) - \Delta_\nu z_\mu(n) + N M_{\mu\nu}(n) ,
 \label{eq:z_1transf}
\end{align}
where we have defined the forward difference,
$\Delta_\mu f(n)\equiv f(n+\Hat\mu)-f(n)$,
and $M_{\mu\nu}(n)$ is required
to restrict the \emph{$2$-form gauge field} $z_{\mu\nu}(n)$ to
\begin{align}
\begin{cases}
 0\leq z_{\mu\nu}(n)<N & \text{for $\mu<\nu$} ,\\
 z_{\mu\nu}(n)=-z_{\nu\mu}(n) & \text{for $\mu>\nu$} .
\end{cases}
\end{align}
Under this $1$-form transformation,
$\Check{P}\mapsto
e^{-\frac{2\pi i}{N}\left(\Delta_\mu z_\nu-\Delta_\nu z_\mu+N M_{\mu\nu}\right)}
\Check{P}$ by our choice of the branch.

As shown in Appendix A of Ref.~\cite{Abe:2022nfq},
the consistency of transition functions among the intersection
of eight hypercubes\footnote{%
In the usual context of the fiber bundle,
the cocycle condition for the transition functions is consistency
at the boundary of three patches;
the flatness condition is at the quadruple overlap~\cite{Tanizaki:2022ngt}.
On the other hand, for the square lattice,
the former is defined in the intersection of four hypercubes;
the latter is in the intersection of eight hypercubes.}
leads us to the flatness of
the $\mathbb{Z}_N$ $2$-form gauge field~$z_{\mu\nu}(n)$.
We can see that $z_{\mu\nu}(n)$ satisfies the modulo~$N$ flatness condition,
\begin{align}
 \frac{1}{2}\sum_{\nu,\rho,\sigma}\epsilon_{\mu\nu\rho\sigma}
 \Delta_\nu z_{\rho\sigma}(n)=0\bmod N .
\end{align}

Therefore, we find that, under the $\mathbb{Z}_N$ $1$-form gauge transformation,
\begin{align}
 \Check{v}_{n,\mu}(x)
 &\mapsto \exp\left\{-\frac{2\pi i}{N}\sum_{\nu>\mu}
 \left[\Delta_\mu z_\nu(n-\Hat\mu)-\Delta_\nu z_\mu(n-\Hat\mu)
 +N M_{\mu\nu}(n-\Hat\mu)\right] y_\nu
 \right\}
 \notag\\&\qquad\times
 \exp\left[\frac{2\pi i}{N}z_\mu(n-\Hat\mu)\right]\Check{v}_{n,\mu}(x) .
\end{align}
The first factor can be canceled against the transformation of~$\omega_{n,\mu}(x)$,
which depends on the gauge field~$z_{\mu\nu}(n)$,
so we have
\begin{align}
 v_{n,\mu}(x) \mapsto \exp\left[\frac{2\pi i}{N}z_\mu(n-\Hat\mu)\right] v_{n,\mu}(x) .
\end{align}
The fractional topological charge $\mathcal{Q}$
\begin{align}
 \mathcal{Q} &= - \frac{1}{8N}\sum_{n\in\Lambda}\sum_{\mu,\nu,\rho,\sigma}
 \epsilon_{\mu\nu\rho\sigma}z_{\mu\nu}(n)z_{\rho\sigma}(n+\Hat\mu+\Hat\nu)
 + \Check{\mathcal{Q}}
\end{align}
is invariant
under the $\mathbb{Z}_N$ $1$-form gauge transformation,
while $\Check{\mathcal{Q}}$ is not.
The shift of $\Check{\mathcal{Q}}$ should vanish
owing to the shift of the first term
(see Appendix A in~Ref.~\cite{Abe:2022nfq}).

\subsection{Reproducing the Kapustin--Seiberg prescription}
To compare some properties between the lattice
and continuum theories~\cite{Kapustin:2014gua},
one may construct the $SU(N)/\mathbb{Z}_N$ principal bundle
from the lattice $U(N)$ gauge theory.
Starting from the above construction by the lattice $SU(N)$ gauge theory,
we perform the $U(1)$ $1$-form transformation given by\footnote{%
This procedure is analogous to the construction
of weak higher-groups from strict higher-groups.
See Ref.~\cite{Kan:2023yhz} for the $U(1)$ gauge theory
or Sect.~\ref{sec:high-group} for the $SU(N)$ gauge theory.}
\begin{align}
 \Check{U}(n,\mu) \mapsto \exp\left[i\lambda_\mu(n)\right] \Check{U}(n,\mu)
 \in U(N)
 \qquad \lambda_\mu(n)\in\mathbb{R},\, 0\leq\lambda_\mu(n)<2\pi .
\end{align}
Now we define the $U(1)$ $2$-form gauge field
\begin{align}
 \lambda_{\mu\nu}(n) \equiv \frac{2\pi}{N}z_{\mu\nu}(n)
 + \Delta_\mu\lambda_\nu(n) - \Delta_\nu\lambda_\mu(n) + 2\pi K_{\mu\nu}(n),
\end{align}
where $K_{\mu\nu}(n)\in\mathbb{Z}$ is
required to restrict $\lambda_{\mu\nu}(n)$ to
\begin{align}
 \begin{cases}
  0\leq\lambda_{\mu\nu}(n)<2\pi & \text{for $\mu<\nu$} ,\\
  \lambda_{\mu\nu}(n) = - \lambda_{\nu\mu}(n) & \text{for $\mu>\nu$}.
 \end{cases}
\end{align}
After successive $U(1)$ $1$-form transformations,
a generic $\lambda_{\mu\nu}(n)$ transforms as
\begin{align}
 \lambda_{\mu\nu}(n) \mapsto \lambda_{\mu\nu}(n)
 + \Delta_\mu\lambda_\nu(n) - \Delta_\nu\lambda_\mu(n) + 2\pi K_{\mu\nu}(n) .
\end{align}

Then, the transition function $v_{n,\mu}(x)\in U(N)$ transforms as
\begin{align}
 v_{n,\mu}(x) \mapsto \exp\left[i\lambda_\mu(n-\Hat\mu)\right] v_{n,\mu}(x) ,
\end{align}
and satisfies the cocycle condition
\begin{align}
 v_{n-\Hat\nu,\mu}(x)v_{n,\nu}(x)v_{n,\mu}(x)^{-1}v_{n-\Hat\mu,\nu}(x)^{-1}
 = \exp\left[i\lambda_{\mu\nu}(n-\Hat\mu-\Hat\nu)\right] .
\end{align}
We have the topological charge
\begin{align}
 \mathcal{Q}
 &= - \frac{N}{32\pi^2}\sum_{n\in\Lambda}\sum_{\mu,\nu,\rho,\sigma}
 \epsilon_{\mu\nu\rho\sigma} \lambda_{\mu\nu}(n)\lambda_{\rho\sigma}(n+\Hat\mu+\Hat\nu)
 + \Check{\mathcal{Q}} .
\end{align}
This is, in fact, $U(1)$ $1$-form invariant,
and thus, takes the same fractional value as before.
This means that we can regard $\lambda_{\mu\nu}(n)$ as a field strength
in the lattice $U(1)/\mathbb{Z}_N$ gauge theory;
as already shown in Ref.~\cite{Abe:2022nfq},
one can show that the first term in~$\mathcal{Q}$ is $N\times\frac{1}{N^2}\mathbb{Z}$.

The field strength is defined by
\begin{align}
 F_{\mu\nu}(n) \equiv \frac{1}{i}\ln\left[P(n,\mu,\nu)\right] ,
\end{align}
and $\Check{F}_{\mu\nu}(n)$ by $\Check{P}(n,\mu,\nu)$,
in the range (or branch) that
those satisfy the admissibility condition~\eqref{eq:admissibility}.
From $P(n,\mu,\nu)= e^{-i\lambda_{\mu\nu}(n)}\Check{P}(n,\mu,\nu)\in SU(N)$,
$F_{\mu\nu}=\Check{F}_{\mu\nu}-\lambda_{\mu\nu}\in \mathfrak{su}(N)$
and $\Check{F}_{\mu\nu}\in\mathfrak{u}(N)$;
we have
\begin{align}
 \Tr\Check{F}_{\mu\nu}(n) = N \lambda_{\mu\nu}(n).
\end{align}
Also under the $U(1)$ $1$-form gauge transformation,
the field strength $F_{\mu\nu}(n)$ is invariant,
and $\Check{F}_{\mu\nu}(n)$ transforms as
\begin{align}
 \Check{F}_{\mu\nu}(n) \mapsto \Check{F}_{\mu\nu}(n)
 + \Delta_\mu\lambda_\nu(n) - \Delta_\nu\lambda_\mu(n) + 2\pi K_{\mu\nu}(n) .
\end{align}
These expressions provide good agreement with those in the continuum theory
by the Kapustin--Seiberg prescription~\cite{Kapustin:2014gua}.
Therefore, our fully-regularized construction can naturally reproduce non-Abelian gauge theories
with discrete higher-form symmetries in the continuum limit.

We make some comments here:
First, in the more sophisticated and minimal construction
given in Ref.~\cite{Abe:2023ncy},
we can define the $1$-form invariant $F_{\mu\nu}(n)$ only,
and so the relation with the coupling with higher-form gauge fields
in the continuum is not obvious.
Also, if we use the $\mathbb{Z}_N$ $2$-form gauge field $z_{\mu\nu}(n)$,
its continuum limit seems to be puzzling
because of no \emph{smooth and integer} fields in the continuum description.
This difficulty reminds us that
there is a certain procedure required
for such a coupling with a $\mathbb{Z}_N$ field in the continuum,
which would be described by a pair of $U(1)$ gauge fields.
We can describe it by the lattice gauge field $\lambda_{\mu\nu}(n)$.

On the other hand, lattice gauge theories can be coupled
quite simply with lattice integer fields,
while there are subtleties in the continuum theory
since cohomological operations are needed
such as the Pontryagin square~\cite{Kapustin:2013qsa,Kapustin:2014gua}.
The Kapusitin--Seiberg prescription gives a simplified explanation
for discrete gauge fields by embedding it to $U(1)$.
Thus, we see the de~Rham cohomology but not the \v{C}ech cohomology;
to circumvent this issue,
the wedge product in expressions as a result
should be replaced by the Pontryagin square.
Such issues make some studies in continuum apparently difficult.

\section{Application: Higher-group structure in modified $SU(N)$ gauge theory}
\label{sec:high-group}
As a simple application,
let us consider the higher-group structure in the $SU(N)$ gauge theory
modified as follows:
The insertion of the delta function in the path integral,
\begin{align}
 \delta(q(n) - p c(n)) ,
\end{align}
restricts the instanton number to integral multiples of~$p\in\mathbb{Z}$
\cite{Pantev:2005rh,Pantev:2005wj,Pantev:2005zs,Seiberg:2010qd}.
Here we have defined $\mathcal{Q}=\sum_{n\in\Lambda}q(n)$,
and $c_{\mu\nu\rho\sigma}(n)$ is the $4$-form field strength
of a compact $U(1)$ $3$-form gauge field
such that\footnote{%
It was proved~\cite{Luscher:1998kn} that there exists the $U(1)$ gauge potential on the lattice
such that $F_{\mu\nu}(n)=\Delta_\mu A_\nu(n)-\Delta_\nu A_\mu(n)$
for admissible gauge configurations.
Here $A_\mu(n)$ is constructed by $a_\mu(n)\equiv\frac{1}{i}\ln U(n,\mu)$
and $F_{\mu\nu}(n)\equiv\frac{1}{i}\ln P(n,\mu,\nu)$,
where $F_{\mu\nu}(n)-\Delta_\mu a_\nu(n)+\Delta_\nu a_\mu(n)\in 2\pi\mathbb{Z}$.
This can be immediately generalized for $U(1)$ higher-form gauge fields.}
\begin{align}
 c_{\mu\nu\rho\sigma}(n) \equiv \epsilon_{\mu\nu\rho\sigma}c(n) ,
\end{align}
and $\sum_{n\in\Lambda}c(n)\in\mathbb{Z}$.
Thus, perturbation theory is not affected at all and the local nature is unchanged,
while globally or topologically speaking this modification possesses
a quite nontrivial structure,
called the higher-group symmetry~\cite{Tanizaki:2019rbk}.\footnote{%
In topological lattice gauge theories,
we can also observe the higher-group structure;
see Refs.~\cite{Delcamp:2018wlb,Delcamp:2019fdp}.}
This theory provides an application of our construction
of the~$SU(N)/\mathbb{Z}_N$ principal bundle.

Note that, to see this, it is important to introduce a technique
called the integral lift~\cite{Kapustin:2013qsa,Kapustin:2014gua},
as mentioned in the case of the lattice $U(1)$ gauge theory
with restricted topological sectors~\cite{Kan:2023yhz}.\footnote{%
Recently, the definition of (higher-)cup products,
from which the Pontryagin square can be constructed,
on the hypercubic lattice is given in~Ref.~\cite{Chen:2021ppt}.
For a torsion-free manifold,
we can also use the integral lift as another possible choice.}
This is because the fractional part of the topological charge~$\mathcal{Q}$
possesses the non-commutativity,
$\sum_{n\in\Lambda}\sum_{\mu,\nu,\rho,\sigma}\epsilon_{\mu\nu\rho\sigma}z_{\mu\nu}(n)z_{\rho\sigma}(n+\Hat\mu+\Hat\nu)$,
and naively seems to become $\frac{1}{2N}$,
but it is $\frac{1}{N}$ thanks to
the commutativity of the original form,
$\sum_{\mu,\nu,\rho,\sigma}\epsilon_{\mu\nu\rho\sigma}z_{\mu\nu}z_{\rho\sigma}$.
Suppose that an integer field $\Bar{z}_{\mu\nu}(n)$
is defined by~$z_{\mu\nu}(n)=\Bar{z}_{\mu\nu}(n)\bmod N$
so that $\sum_{\nu,\rho,\sigma}\epsilon_{\mu\nu\rho\sigma}
\Delta_\nu\Bar{z}_{\rho\sigma}(n)=0$.
Then, replacing $z_{\mu\nu}(n)$ by~$\Bar{z}_{\mu\nu}(n)$,
we have the desired $\mathcal{Q}\in\frac{1}{N}\mathbb{Z}$ for a generic~$\Bar{z}_{\mu\nu}(n)$.

Again, let us consider the constraint,
\begin{align}
 q(n) - p c(n) = 0 .
\end{align}
At first sight, any nontrivial configuration of $\Bar{z}_{\mu\nu}(n)$ is forbidden,
so gauging the $\mathbb{Z}_N$ $1$-form symmetry seems to be impossible.
Now, by introducing the $\mathbb{Z}_p$ $3$-form symmetry
and gauging these two symmetries simultaneously,
we have a nontrivial theory with the $4$-group structure.
This can be immediately realized by the replacement
as~$c\to c-\frac{1}{N p}\Bar{w}$, where a $4$-form field $\Bar{w}(n)\in\mathbb{Z}$;
then
\begin{align}
 q(n) - p c(n) + \frac{1}{N}\Bar{w}(n) = 0.
\end{align}
Because all fractional contributions from~$q(n)$ by~$\Bar{z}_{\mu\nu}(x)$
can be absorbed into~$\Bar{w}(n)$,
one can obtain nontrivial configurations of~$\Bar{z}_{\mu\nu}(n)$.
Note that, by construction, we see the $3$-form gauge symmetry,
\begin{align}
 \Bar{w}(n) &\mapsto \Bar{w}(n)
 + \frac{1}{3!}
 \sum_{\mu,\nu,\rho,\sigma}\epsilon_{\mu\nu\rho\sigma}\Delta_\mu w_{\nu\rho\sigma}(n)
 + N p\Bar{M}(n) ,\label{eq:3_transf_w}\\
 c(n) &\mapsto c(n) + \frac{1}{N p}
 \frac{1}{3!}
 \sum_{\mu,\nu,\rho,\sigma}\epsilon_{\mu\nu\rho\sigma}\Delta_\mu w_{\nu\rho\sigma}(n)
 + \Bar{M}(n) .
\end{align}
Here we assume that $w_{\mu\nu\lambda}(n)\in\mathbb{Z}$ satisfies
$\frac{1}{3!}\sum_{n\in\Lambda}\sum_{\mu,\nu,\rho,\sigma}
\Delta_\mu w_{\nu\rho\sigma}(n)\in N p\mathbb{Z}$ and
$0\leq w_{\mu\nu\lambda}(n)<N p$ for~$\mu<\nu<\lambda$;
$\Bar{M}(n)$ is an integer field.

Now, we define $w(n)\in\mathbb{Z}$ as~$w(n)=\Bar{w}(n)\bmod N$ and $0\leq w(n)<N p$.
A new field $M(n)\in\mathbb{Z}$ instead of~$\Bar{M}(n)$ is introduced
for $w(n)$ to be $0\leq w(n)<N p$
under the $3$-form gauge transformation~\eqref{eq:3_transf_w} for~$w(n)$.
That is, $\Bar{w}(n)$ is the integral lift of~$w(n)$.
$w(n)$ should be regarded as the $\mathbb{Z}_{N p}$ $4$-form gauge field,
while the original $3$-form symmetry is $\mathbb{Z}_p$
if one ignore $\Bar{z}_{\mu\nu}(n)$ and~$N$.
Thus, this is the (strict) $4$-group structure.

As another perspective,
we can consider the use of~$\Omega(n)\in\mathbb{R}$
such that $c\to c-\frac{1}{N p}\Omega$.
Then, we have
\begin{align}
 q(n) - p c(n) + \frac{1}{N}\Omega(n) = 0 ,
\end{align}
where we assume that $\sum_{n\in\Lambda}\Omega(n)\in\mathbb{Z}$
to remove the fractionality in the first term.
We now define $\Tilde\Omega(n)\in\mathbb{R}$ by
\begin{align}
 \Tilde\Omega(n)
 &\equiv \frac{1}{N}\Omega(n)
 - \frac{1}{8N}\sum_{\mu,\nu,\rho,\sigma}\epsilon_{\mu\nu\rho\sigma}
 \Bar{z}_{\mu\nu}(n)\Bar{z}_{\rho\sigma}(n+\Hat\mu+\Hat\nu) ,
 \label{eq:tilde_omega}
\end{align}
where $\sum_{n\in\Lambda}\Tilde\Omega(n)\in\mathbb{Z}$.
Therefore, by using~$\Check{\mathcal{Q}}=\sum_{n\in\Lambda}\Check{q}(n)$,
we find that the constraint becomes
\begin{align}
 \Check{q}(n) - p c(n) + \Tilde\Omega(n) = 0.
\end{align}
From the definition of~$\Tilde\Omega(n)$ \eqref{eq:tilde_omega},
$\Tilde\Omega(n)$ is not invariant any longer
under the $\mathbb{Z}_N$ $1$-form gauge transformation,
while $[\Check{q}(n)+\Tilde\Omega(n)]$ is invariant.
One can find that, owing to the integral lift,
$\sum_{n\in\Lambda}\Tilde\Omega(n)\in\mathbb{Z}$ holds
after we perform the $1$-form gauge transformation.

Following the procedure given in Ref.~\cite{Kan:2023yhz},
we can compel $\Tilde\Omega(n)\in\mathbb{R}$
to become~$\Tilde{\Bar{w}}(n)\in\mathbb{Z}$
which is the integral lift as~$\Tilde{w}(n)=\Tilde{\Bar{w}}(n)\bmod p$.
This is always possible since we can throw away the real or fractional part apart from integers
in~$\Tilde\Omega(n)$ into the gauge redundancy of~$c(n)$,
by using the \emph{continuum} $3$-form transformation,
\begin{align}
 \Tilde\Omega(n) &\mapsto \Tilde\Omega(n)
 + \frac{1}{3!}\sum_{\mu,\nu,\rho,\sigma}\epsilon_{\mu\nu\rho\sigma}
 \Delta_\mu\Tilde\Omega_{\nu\rho\sigma}(n) + p\Tilde{M}(n) ,\\
 c(n) &\mapsto c(n)
 + \frac{1}{p} \frac{1}{3!}\sum_{\mu,\nu,\rho,\sigma}\epsilon_{\mu\nu\rho\sigma}
 \Delta_\mu\Tilde\Omega_{\nu\rho\sigma}(n) + \Tilde{M}(n) ,
\end{align}
where $\Tilde\Omega_{\mu\nu\lambda}(n)\in\mathbb{R}$ and $\Tilde{M}(n)\in\mathbb{Z}$.
The ``$\mathbb{Z}_N$ $1$-form gauge transformation''
is redefined as the original $1$-form gauge transformation
and such a $3$-form transformation to be set to integers.
This theory possesses the modified ``$1$-form''
and the \emph{discrete} $\mathbb{Z}_p$ $3$-form gauge symmetries,
that is, the (weak) $4$-group symmetry.

\section{Conclusion}
We constructed the $SU(N)/\mathbb{Z}_N$ principal bundle
from the $SU(N)$ lattice gauge theory with the 't~Hooft twisted boundary condition.
This construction requires the appropriate admissibility,
which is ensured by the proof based on the principle of the locality,
$SU(N)$ gauge invariance,
and the $\mathbb{Z}_N$ $1$-form gauge invariance~\cite{Abe:2023ncy}.
We provided the concrete expressions for
not only the twisted variables [e.g., $v_{n,\mu}(x)$],
which can be equivalently described by those in Ref.~\cite{Abe:2023ncy},
but also the periodic variables [e.g., $\Check{v}_{n,\mu}(x)$]
and the $\mathbb{Z}_N$ $1$-form gauge field $z_{\mu\nu}(n)$.
In our construction, the periodic variables enjoy the structure of the $U(N)$ principal bundle
rather than $SU(N)$ as the continuum theory.
This fact thus leads us to explicitly reproduce the Kapustin--Seiberg prescription
in terms of the lattice fields,
and quite naturally depict its behavior in the continuum limit.
Also, similarly to Ref.~\cite{Abe:2023ncy},
we can observe the mixed 't~Hooft anomaly for the $\mathbb{Z}_N$ $1$-form gauge symmetry and the $\theta$ periodicity,
and so on.

We further considered the instanton-sum modified $SU(N)$ lattice gauge theory.
It was first shown that naively gauging the $\mathbb{Z}_N$ $1$-form symmetry
is impossible under the constraint of the instanton numbers restricted to~$p\mathbb{Z}$.
As the strict $4$-group,
we introduced the $\mathbb{Z}_{N p}$ $4$-form gauge field~$w(n)\in\mathbb{Z}$
associated with the $\mathbb{Z}_{N p}$ $3$-form gauge symmetry to compensate this difficulty.
Also we showed the weak $4$-group (Green--Schwarz-type) structure
such that the $\mathbb{Z}_p$ $4$-form gauge field~$\Tilde{w}(n)\in\mathbb{Z}$
transforms not only under the discrete $\mathbb{Z}_p$ $3$-form gauge symmetry,
but also under the ``mixed $1$-form'' gauge symmetry
that includes the original $\mathbb{Z}_N$ $1$-form
and the continuum $\mathbb{Z}_p$ $3$-form gauge symmetries.

The non-invertible symmetry is still developing,
and so we hope to apply our approach to such recent developments.
It is also interesting to consider the case with matter fields,
especially lattice fermions,
whose construction is quite nontrivial in lattice gauge theory.
Also the index theorem on the lattice is an attractive issue
with the consideration of an appropriate overlap Dirac operator.

\section*{Acknowledgements}
We would like to especially thank
Hiroshi Suzuki and Yuya Tanizaki for useful discussions and collaboration.
The work of M.A. was supported by Kyushu University Innovator Fellowship Program in Quantum Science Area.
O.M. is grateful to Naoto Kan, Yuta Nagoya, and Hiroki Wada for helpful discussions.
This work of O.M. was supported by Japan Society for the Promotion of Science (JSPS)
Grant-in-Aid for Scientific Research Grant Number JP21J30003.

\appendix
\section{L\"uscher's interpolation functions}
\label{sec:luscher_formulas}
We summarize the L\"uscher's construction for the transition function~\cite{Luscher:1981zq,Abe:2023ncy}.
First, we define new link variables
which are constructed, in terms of the parallel transport functions
$w^n(x)$~\eqref{eq:parallel_trans}, by
\begin{align}
 \begin{cases}
  u^n_{x y}=w^n(x)U(x,\mu)w^n(y) & \text{if $y=x+\Hat{\mu}$} ,\\
  u^n_{x y}=(u^n_{yx})^{-1} & \text{if $y=x-\Hat{\mu}$} .
 \end{cases}
\end{align}
By these link variables, we can construct the transition function
at the corners of~$f(n,\mu)$,
\begin{align}
    u^{n-\Hat{\mu}}_{xy}&=v_{n,\mu}(x)u^n_{xy}v_{n,\mu}(y)^{-1},\\
    v_{n,\mu}&\equiv w^{n-\Hat{\mu}}(x)w^{n}(x)^{-1}.
\end{align}
 Next, in order to calculate the topological charge~\eqref{eq:topo_charge_lat},
 we interpolate the transition function to $\forall x\in f(n,\mu)$.
 Then, we need to define the interpolation function~$S^m_{n,\mu}$.
 We first label the corners $\{s_i\}_{i=0,\dots,7}$ of $f(n,\mu)$ as follows:
 letting $\alpha$, $\beta$, $\gamma\in\{1,2,3,4\}\setminus\{\mu\}$
 and $\alpha<\beta<\gamma$,
 \begin{align}
     s_0&=n,& s_1&=n+\Hat\alpha,& s_2&=n+\Hat\beta,& s_3&=n+\Hat\gamma,\notag\\
     s_4&=n+\Hat\alpha+\Hat\beta+\Hat\gamma,& s_5&=n+\Hat\alpha+\Hat\gamma,&
     s_6&=n+\Hat\alpha+\Hat\beta,& s_7&=n+\Hat\beta+\Hat\gamma.
\end{align}
Then, for $m=n$, $n-\hat{\mu}$, we define the interpolation functions,
\begin{align}
  f^m_{n,\mu}(x_\gamma)
  &=(u^m_{s_3s_0})^{y_\gamma}(u^m_{s_0s_3}u^m_{s_3s_7}u^m_{s_7s_2}
  u^m_{s_2s_0})^{y_\gamma}u^m_{s_0s_2}(u^m_{s_2s_7})^{y_\gamma},\\
  g^m_{n,\mu}(x_\gamma)
  &=(u^m_{s_5s_1})^{y_\gamma}(u^m_{s_1s_5}u^m_{s_5s_4}u^m_{s_4s_6}
  u^m_{s_6s_1})^{y_\gamma}u^m_{s_1s_6}(u^m_{s_6s_4})^{y_\gamma},\\
  h^m_{n,\mu}(x_\gamma)
  &=(u^m_{s_3s_0})^{y_\gamma}(u^m_{s_0s_3}u^m_{s_3s_5}u^m_{s_5s_1}
  u^m_{s_1s_0})^{y_\gamma}u^m_{s_0s_1}(u^m_{s_1s_5})^{y_\gamma},\\
  k^m_{n,\mu}(x_\gamma)
  &=(u^m_{s_7s_2})^{y_\gamma}(u^m_{s_2s_7}u^m_{s_7s_4}u^m_{s_4s_6}
  u^m_{s_6s_2})^{y_\gamma}u^m_{s_2s_6}(u^m_{s_6s_4})^{y_\gamma},\\
  l^m_{n,\mu}(x_\beta,x_\gamma)
  &=\left[ f^m_{n,\mu}(x_\gamma)^{-1} \right]^{y_\beta}
  \left[ f^m_{n,\mu}(x_\gamma)k^m_{n,\mu}(x_\gamma)g^m_{n,\mu}
  (x_\gamma)^{-1}h^m_{n,\mu}(x_\gamma)^{-1} \right]^{y_\beta}\notag\\&\qquad
  \cdot h^m_{n,\mu}(x_\gamma)\left[ g^m_{n,\mu}(x_\gamma) \right]^{y_\beta},\\
  S^m_{n,\mu}(x_\alpha,x_\beta,x_\gamma)
  &=(u_{s_0s_3}^m)^{y_\gamma}
  \left[ f^m_{n,\mu}(x_\gamma) \right]^{y_\beta}
  \left[ l^m_{n,\mu}(x_\beta,x_\gamma) \right]^{y_\alpha}.
\end{align}
Here, $y_{\lambda}\equiv x_{\lambda}-n_{\lambda}$
and $0\le y_{\lambda}\le1$ for $\lambda=\alpha$, $\beta$, $\gamma$.
We have constructed $S_{n,\mu}^m(x_\alpha,x_\beta,x_\gamma)$
based on the link variable~$U(n,\mu)$;
we simply write it as
$S_{n,\mu}^m(U;x)=S_{n,\mu}^m(x_\alpha,x_\beta,x_\gamma)[U]$ in Sect.~\ref{sec:3.2}.

\bibliographystyle{utphys}
\bibliography{ref}
\end{document}